\documentclass[aps,prl,twocolumn,groupedaddress,showpacs]{revtex4}

\usepackage{color}
\usepackage{graphicx}
\usepackage{ulem}
\usepackage{bm}
\newcommand{\tcr}{\textcolor{black}}

\begin{document}



\title{Derivation of the spin-glass order parameter from stochastic thermodynamics}


\author{A. Crisanti$^1$}
\author{M. Picco$^2$}
\author{F. Ritort$^3$}

\affiliation{$^1$Dipartimento di Fisica, Universit\`a di Roma
             ``La Sapienza'', and \\ 
            Istituto dei Sistemi Complessi ISC-CNR, 
            P.le Aldo Moro 2, 00185 Roma, Italy}

\affiliation{$^2$Sorbonne Universit\'e, CNRS, Universit\'e Pierre et Marie Curie, LPTHE, F-75005 Paris, France}


\affiliation{$^3$Small Biosystems Lab, Condensed Matter Physics Department, Universitat de Barcelona, C/ Mart\'{\i} i Franqu\`es, E-08028, Barcelona}
\affiliation{CIBER-BBN Center for Bioengineering, Biomaterials and Nanomedicine, Instituto de Salud Carlos III, Madrid}

\date{\today}



\begin{abstract}
\noindent A fluctuation relation is derived to extract the order parameter function $q(x)$ in \tcr{weakly ergodic} systems. The relation is based on measuring and classifying entropy production fluctuations according to the value of the overlap $q$ between configurations. For a fixed value of $q$, entropy production fluctuations are Gaussian distributed allowing us to derive the quasi-FDT so characteristic of aging systems. The theory is validated by extracting the $q(x)$ in various types of glassy models. It might be generally applicable to other \tcr{nonequilibrium systems} and experimental small systems.
\end{abstract}


\maketitle

\begin{figure}[h] 
   \centering
   \includegraphics[width=2.5in]{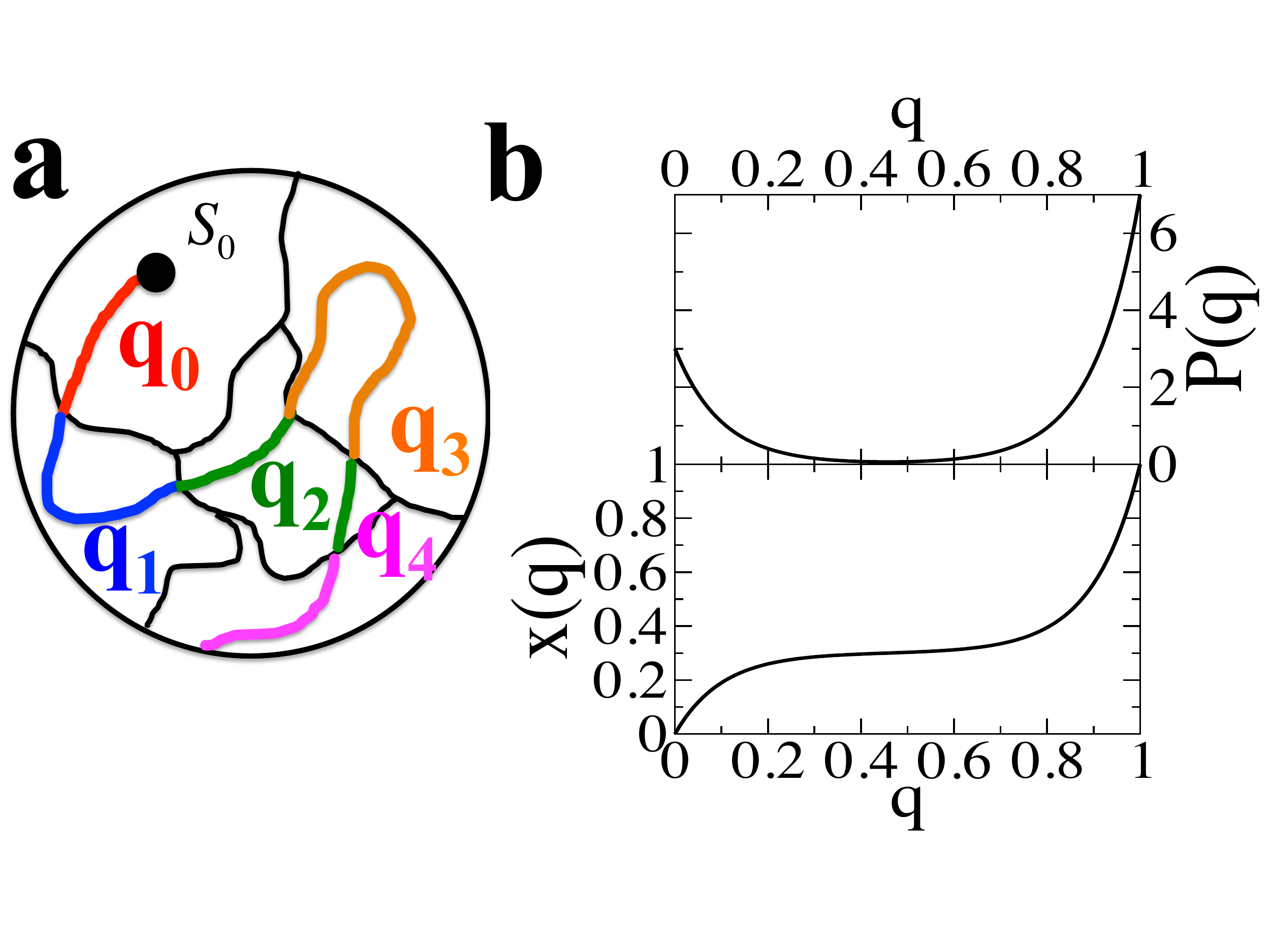} 
   \caption{{\bf The order parameter $q(x)$.} Illustration of a path in phase space (a). The time evolution of correlations $q_i=C(S_0,S_i)$ between different states $S_i$ and state $S_0$ at $t_w$ is collected to build a histogram $P(q)=\langle\delta(q-q_i)\rangle$ where $<..>$ denotes the average over paths (panel b, upper). $x(q)$ is related to $P(q)$, $P(q)=dx(q)/dq$ (panel b, lower).}
   \label{fig:FIG1}
\end{figure}

Dynamically frustrated systems (DFS) include a wide category of soft matter and solid state systems (e.g. polymers, gels, superconductors, magnetic alloys, biomolecules, etc..)  exhibiting strong nonequilibrium effects characterized by slow
relaxational kinetics, aging and memory effects \cite{SG,Cav09,BerBir11}. 
 Such a rich
phenomenology has been predicted by  weak ergodicity breaking models, yet resists interpretation in terms of equilibrium-based statistical thermodynamics approaches.  Experimentally, a spatial
correlation length mildly growing in time spanning a few intermolecular distances has been
identified in DFS suggesting that some kind of ordering takes place \cite{BerBirBouCipMasHotLadPie05}. The order parameter in DFS, originally introduced in the context of spin glasses as a
time-dependent correlation function \cite{EdwAnd75}, still eludes direct experimental measurement. In
contrast to homogeneous systems where one or a few single order
parameters quantify the degree of macroscopic order, in DFS a whole
function $q(x)$ is required. Roughly speaking, $q(x)$ quantifies
the degree of correlation $q$ between different states that share a
common fraction $x$ of thermalized degrees of freedom in a corrugated
free energy landscape with many states separated by high free energy
barriers \cite{Par83}. The mathematical description of such landscape is often known as the
replica symmetry breaking solution of the mean-field spin glass \cite{MezParVir88}.  A key aspect of $q(x)$ is the possibility to define it from
statics and dynamics reflecting the fact that some kind of
thermalization, characteristic of equilibrium systems, still operates
under weakly ergodic conditions \cite{FraVir00}.  In fact, DFS partially equilibrate among the 
many states comprised
by the $q(x)$, however they do so only for finite observation times as they intermittently change state during the relaxation process. For
systems with a non-trivial $q(x)$, its inverse function $x(q)$ has been interpreted in terms of frozen, i.e., not-relaxed, degrees of freedom \cite{CugKurPel97,CriDom11} with $P(q)=dx(q)/dq$ positively
defined. In spin-glass theory $P(q)$ is defined as the probability of
states having overlap $q$, $dx$ being the probability
measure for the states. To extract the equilibrium $x(q)$ requires monitoring the overlap between the state of the system at time $t$ and an initially equilibrated state and calculate the overlap distribution $P(q)$ (Fig.\ref{fig:FIG1}). However, in DFS such measurement is hard because one cannot equilibrate nor efficiently sample the different states.

A different route to extract $q(x)$
is to use a modified version of the fluctuation-dissipation theorem (referred to as quasi-FDT \cite{CriHorSom93,CugKur93,Cug03,CriRit03}) to measure
time-dependent correlations and responses and extracting the so-called fluctuation dissipation ratio (FDR) $\tilde{x}(q)$ (Eq.(\ref{QFDT}) below). The dynamical FDR $\tilde{x}(q)$ has been shown to be equal to the $x(q)$ from statics \cite{FraMezParPel98}, however its experimental measurement remains difficult \cite{HerOci02}. Here we present a novel and
easier approach to measure $q(x)$ in weakly ergodic systems based on
noise measurements. A first step in this direction was already
presented by us a few years ago \cite{CriPicRit13} by deriving a
fluctuation relation (FR) for the aging state under some
assumptions. \tcr{Here we make a step further and introduce a FR for entropy production fluctuations at fixed overlap $q$ and use it to extract the order parameter $x(q)$ in DFS under weak ergodicity breaking conditions. \iffalse The new FR provides a natural connection with the quasi-FDT.\fi}

\tcr{The key idea of the approach is to measure the entropy
produced by an external perturbation of strength $h$ that is
applied to the system at a given time $t_w$ during the relaxational
process (e.g. following a T-quench). The field
$h$ is conjugated to the macroscopic observable $M$ that is
experimentally accessible by following its time evolution, $M(t)$, up to
a maximal time. The
experiment is then repeated many times and the probability
distribution $P_{t_w,q}(\Delta S)=\langle \delta\bigl(\Delta S-\Delta S_{t_w,t} \bigr)\rangle_{q=C(t,t_w)}$ with $\Delta S_{t_w,t}=h\Delta M/T$ and $\Delta M=M(t)-M(t_w)$
measured. The average $\langle ... \rangle_{q=C(t,t_w)}$ is constrained to those paths with
$C(t,t_w)=q$ for $t>t_w$, an average that we will refer to as $q$-statistics (or simply $q$-stat). In Ref.\cite{CriPicRit13} we proved that  the unconstrained distribution $P_{t_w}(\Delta S)$, defined as the probability of observing the value $\Delta S_{t_w,t}$ after $t_w$ (i.e. without classifying states) satisfies a FR with a large deviation parameter $x_{t_w}$. In the $q$-stat scheme we can introduce the large deviation function $x_{t_w}(q)$ through the following FR,
\begin{equation}
\label{eqLDF2}
x_{t_w}(q)=\frac{k_B}{\Delta S}\log\Bigl(\frac{P_{t_w,q}(\Delta S)}{P_{t_w,q}(-\Delta S)} \Bigr)
\end{equation}
with $k_B$ the Boltzmann constant. In a weak ergodicity breaking scenario $q$ is the appropriate parameter to partition phase space into partially equilibrated regions. We hypothesize that $P_{t_w,q}(\Delta S)$ is Gaussian for a system with a sufficiently large number of degrees of freedom.  We will call this the Gaussian Universality Hypothesis (GUH) for $P_{t_w,q}(\Delta S)$. Its simplicity resonates with its importance:  GUH provides an operational definition of partially equilibrated states in nonequilibrium weakly ergodic systems. Its range of validity might go beyond DFS, from non-interacting  systems to nonequilibrium steady states. Let us note that the validity of the GUH is restricted to $q$-stat classification and should not hold for other types of classification among paths, such as fixing the time $t$ after $t_w$ at which $\Delta S_{t_w,t}$ is measured (referred to as $t$-stat) or in the procedure of Ref.\cite{CriPicRit13} (where no classification is made). In fact, in these latter cases the intermittent dynamics mixes states with different values of $q$ and $x(q)$ and  partial equilibration does not hold anymore. In what follows we show that such $q$-stat classification of partially equilibrated states and the GUH imply the validity of the quasi-FDT in DFS, Eq.(\ref{QFDT}). }

Equation (\ref{eqLDF2}) has a simple solution for a Gaussian $P_{t_w,q}(\Delta
S)$ of mean $\langle \Delta S \rangle$ and variance $\sigma^2_{\Delta
  S}$,
\begin{equation}
\label{xqGaussian}
x_{t_w}(q)=\frac{2k_B\langle \Delta S \rangle}{\sigma^2_{\Delta
  S}}\,\,.
\end{equation} 
The advantage of (\ref{xqGaussian}) with respect to (\ref{eqLDF2}) is clear. In order to measure $x_{t_w}(q)$ one does not need to
sample the leftmost tails of $P_{t_w,q}(\Delta
S)$ (i.e. the $\Delta S<0$ events) but just the
mean and variance of the distribution, a more amenable task in experiments and simulations. 

From (\ref{xqGaussian}) we now derive the quasi-FDT. In the linear
response regime we have, 
\begin{equation}
\label{DS1}
\langle \Delta S\rangle=\frac{h\langle \Delta M \rangle}{T}=\frac{h}{T}\Bigl (h\int_{t_w}^t dsG(t,s)+{\cal
  O}(h^3)\Bigr)
\end{equation}  
where we have used the response function definition, $G(t,s)=\frac{\delta \langle M(t)\rangle }{\delta h(s)}$. Moreover,
\begin{eqnarray}
  \sigma^2_{\Delta S}=\Bigl(
\frac{h}{T}\Bigr)^2\bigl(\tilde{C}(t,t)+\tilde{C}(t_w,t_w)-2\tilde{C}(t,t_w)\bigr)=\nonumber\\
\Bigl(\frac{h}{T}\Bigr)^2\bigl(\tilde{C}_0(t,t)+\tilde{C}_0(t_w,t_w)-2\tilde{C}_0(t,t_w)\bigr)+{\cal
O}(h^2)
\label{DS2}
\end{eqnarray} 
with $\tilde{C}(t,s)=\langle M(t)M(s)\rangle$ the two-times
correlation function of the macroscopic observable $M$ in the presence
of a field $h$ (the subscript 0 denotes the same correlation at
$h=0$). Substituting Eqs.(\ref{DS1},\ref{DS2}) in (\ref{xqGaussian}) we
obtain,
\begin{equation}
x_{t_w}(q)=\frac{2k_BT\int_{t_w}^t dsG(t,s)}{\tilde{C}_0(t,t)+\tilde{C}_0(t_w,t_w)-2\tilde{C}_0(t,t_w)}+{\cal
O}(h^2)\,.
\label{xtwq1}
\end{equation}
Let us now consider the quasi-FDT given by \cite{Cug03,CriRit03},
\begin{equation}
k_BTG(t,s)=\theta(t-s)\tilde{x}[\tilde{C}_0(t,s)]\partial_s\tilde{C}_0(t,s)
\label{QFDT}
\end{equation}
with $\tilde{x}[\tilde{C}_0]$ the fluctuation-dissipation ratio
(FDR) which depends on times $t,s$ only through the value of $\tilde{C}_0(t,s)$. Next we derive a relation between   $\tilde{x}(\tilde{C}_0)$ (\ref{QFDT}) and $x_{t_w}(q)$ (\ref{eqLDF2}). Inserting (\ref{QFDT}) in
(\ref{xtwq1}) we get to first order in $h$,
\begin{equation}
\label{xtwq2}
x_{t_w}(q)=\frac{2\int_{\tilde{C}_0(t,t_w)}^{\tilde{C}_0(t,t)}
d\tilde{q}\tilde{x}(\tilde{q})}{\tilde{C}_0(t,t)+\tilde{C}_0(t_w,t_w)-2\tilde{C}_0(t,t_w)}+{\cal
O}(h^2)
\end{equation}
with $\tilde{x}(\tilde{q}=\tilde{C}_0)$ the FDR in
(\ref{QFDT}). If we now assume
$\tilde{C}_0(t_w,t_w)=\tilde{C}_0(t,t)=N$ (e.g. for Ising-type systems with $N$ the total number
of spins) and define
$C(t,s)=\tilde{C}(t,s)/\tilde{C}(s,s)=\tilde{C}(t,s)/N$ we have
$q=C_0(t,t_w)$ and,
\begin{equation}
\label{xtwq3}
x_{t_w}(q)=\frac{1}{1-q}\int_{q}^{1}
dq'\tilde{x}(q')+{\cal
O}(h^2)\,.
\end{equation}
By taking the derivative of (\ref{xtwq3}) with respect to $q$ simple algebra leads to
\begin{equation}
\label{xtwq4}
\tilde{x}(q)=x_{t_w}(q)-(1-q)\frac{dx_{t_w}(q)}{dq} +{\cal O}(h^2)\,.
\end{equation}
directly relating  $\tilde{x}(q)$ and $x_{t_w}(q)$. Note that (\ref{xtwq3}) can be rewritten as $\int_q^1 dq'\tilde{x}(q')=x_{t_w}(q)(1-q)+{\cal O}(h^2)$ meaning that, if we neglect ${\cal O}(h^2)$ corrections, $\tilde{x}(q)$ equals the local slope of the curve $x_{t_w}(q)(1-q)$ versus $1-q$. 

Equation (\ref{xqGaussian}) provides a way to directly extract the static quantities $x(q)$ and $q(x)$ solely from noise $\Delta M=M(t)-M(t_w)$ measurements.  The easiest procedure is to collect statistics of the stochastic variable $\Delta S=h \Delta M/T$ for a fixed value of $q$, apply (\ref{xqGaussian}) and extract the local slope of the curve $x_{t_w}(q)(1-q)$ versus $1-q$. Note that this is the equivalent of the quasi-FDT representation (\ref{xtwq1}) where one plots the susceptibility versus the correlation \cite{Cug03,CriRit03}. Alternatively, one can calculate the derivative of the average entropy production with respect to its variance,
\begin{equation}
\label{xtwq5}
x(q)=2k_B\frac{\partial\langle \Delta S \rangle}{\partial \sigma^2_{\Delta
  S}}=\frac{2k_BT}{h}\frac{\partial\langle \Delta M \rangle}{\partial \sigma^2_{\Delta
  M}}\,.
\end{equation}
%

In order to test the validity of the theory we carried out extensive numerical simulations of three types of DFS. The first example is the Random Orthogonal Model (ROM), a mean-field spin-glass model with one-step replica symmetry breaking (1RSB), defined by \cite{CriRit03},
\begin{equation}
\label{ROM}
{\cal H}=-\sum_{1\leq i < j\leq N}J_{ij}\sigma_i\sigma_j
\end{equation}
where $\sigma_i=\pm 1$ are Ising spins and $J_{ij}=J_{ji}$ are quenched
Gaussian variables of zero mean and variance $1/N$ satisfying $\sum_{k}J_{ik} J_{kj}=16\delta_{ij}$, with $J_{ii}=0$. The system is
perturbed by applying a uniform magnetic field of strength $h$ conjugated to the total magnetization $M(t) = \sum_i \sigma_i(t)$. 

 \begin{figure}[h] 
   \centering
      \includegraphics[width=3.3in]{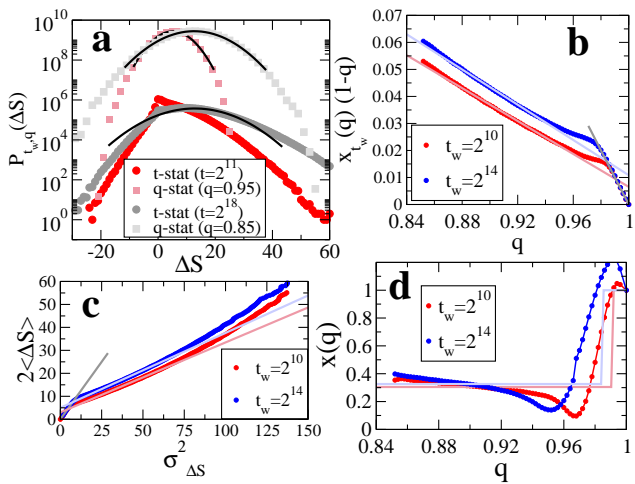} 
   \caption{{\bf Numerical test for ROM and fit to the 1RSB model.} (a) $q$-stat and $t$-stat distributions for $t_w=2^{10}$ and values of $t$ and $q$ such that $C(t,t_w)\simeq q$. Cases are: $C(t=2^ {11},t_w)=0.95$ (red), $C(t=2^ {18},t_w)=0.85$ (grey). Black lines are Gaussian fits that are good only for $q$-stat. (b) $(1-q)x_{t_w}(q)$ versus $q$ from (\ref{xqGaussian}). Results have been fit to the 1RSB model: $(1-q)x_{t_w}(q)=1-q$ for $q\ge q_{EA}$ (grey line) and $1-q_{EA}+m(q_{EA}-q)$ for $q\le q_{EA}$ with $q_{EA}$ and $m$ fitting parameters. We get $q_{EA}=0.99, m=0.30$ ($t_w=2^{10}$, light red) and $q_{EA}=0.98, m=0.33$ ($t_w=2^{14}$, light blue). (c) Alternative method based on (\ref{xtwq5}) with fits taken from panel b. (d) $x(q)$ obtained from the results in panel b using (\ref{xtwq4}) and compared to the 1RSB model (light color lines). c) and d) same color code as in panel b.  Simulations with $T=0.2,N=1000,h=0.1$ and 10M quenches with one sample per quench.}
   \label{fig:FIG2}
\end{figure}

The ROM has a dynamical transition at $T_d=0.536$ where the order parameter jumps from 0 above $T_d$ to $q_{EA}(T_d)=0.96$. Below $T_d$ the system is dynamically confined into one of the exponentially many metastable states with $q(x)=q_{EA}(T)\theta(x-m(T))$. Both $q_{EA}(T)$ and $m(T)$ monotonically decrease with $T$ indicating the gradual freezing of degrees of freedom as $T\to 0$.  Results for the ROM are shown in Figure~\ref{fig:FIG2} after quenching the system from $T=\infty$ to $T=0.2$ for two values of $t_w$. As shown in Fig.\ref{fig:FIG2}a $P(\Delta S)$ for $q$-stat are Gaussian whereas they are not for $t$-stat  \cite{CriPicRit13}. Fig.\ref{fig:FIG2}b plots $x_{t_w}(q)(1-q)$ versus $1-q$ (from (\ref{xqGaussian})) fitted to the theoretical prediction (\ref{xtwq3}) with $\tilde{x}(q)$ the inverse of $q(x)=q_{EA}(T)\theta(x-m(T))$, i.e. $x(q\le q_{EA})=m;  x(q>q_{EA})=1$. Fig.2c shows the construction based on (\ref{xtwq5}) where each point in the plot was obtained by extracting the mean and variance of $P_{t_w,q}(\Delta M)$ for a given value of $q$. Finally, Fig.2d shows the $x(q)$ obtained by any of the two methods. 

\begin{figure}[h] 
   \centering
   \includegraphics[width=3.3in]{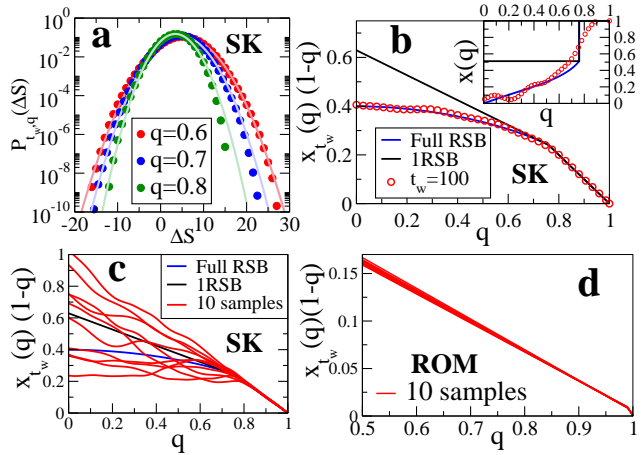} 
   \caption{{\bf Numerical test in the SK model and self-averaging behavior.} (a) $q$-stat distributions for $t_w=100$ for three values of $q$ such that $C(t,t_w)\simeq q$. Light-colour lines are Gaussian fits. (b) $(1-q)x_{t_w}(q)$ and $x(q)$ (inset) versus $q$ from (\ref{xqGaussian}) for $t_w=100$ (red) averaged over 10 samples. Results are compared to exact 1RSB and full RSB models \cite{CriRiz02} showing excellent agreement with the latter. (c)  The same as in (b) but for the 10 samples showing large sample-to-sample fluctuations. (d) As a comparison we show the same as in (c) for the ROM where sample-to-sample fluctuations are small. Details of the fit are given in the caption of Fig.\ref{fig:FIG2}b. Simulations for SK  were done with $T=0.3,N=300,h=0.1$.}
   \label{fig:FIG3}
\end{figure}

Our second example is the Sherrington-Kirkpatrick (SK) mean-field model described by (\ref{ROM}) with uncorrelated Gaussian couplings of zero mean and variance $1/N$, which exhibits full RSB. Fig.\ref{fig:FIG3}a shows again that $P(\Delta S)$ for $q$-stat are Gaussian. From (\ref{xqGaussian}) we extract the $x_{t_w}(q)(1-q)$ which averaged over 10 samples excellently fit the full RSB theoretical prediction \cite{CriRiz02} (Fig.\ref{fig:FIG3}b, main and inset). A main feature of spin-glass theory is the presence of sample-to sample fluctuations which are large for SK (Fig.(\ref{fig:FIG3}c) as compared to ROM (Fig.\ref{fig:FIG3}d). 

\begin{figure}[h] 
   \centering
   \includegraphics[width=3.3in]{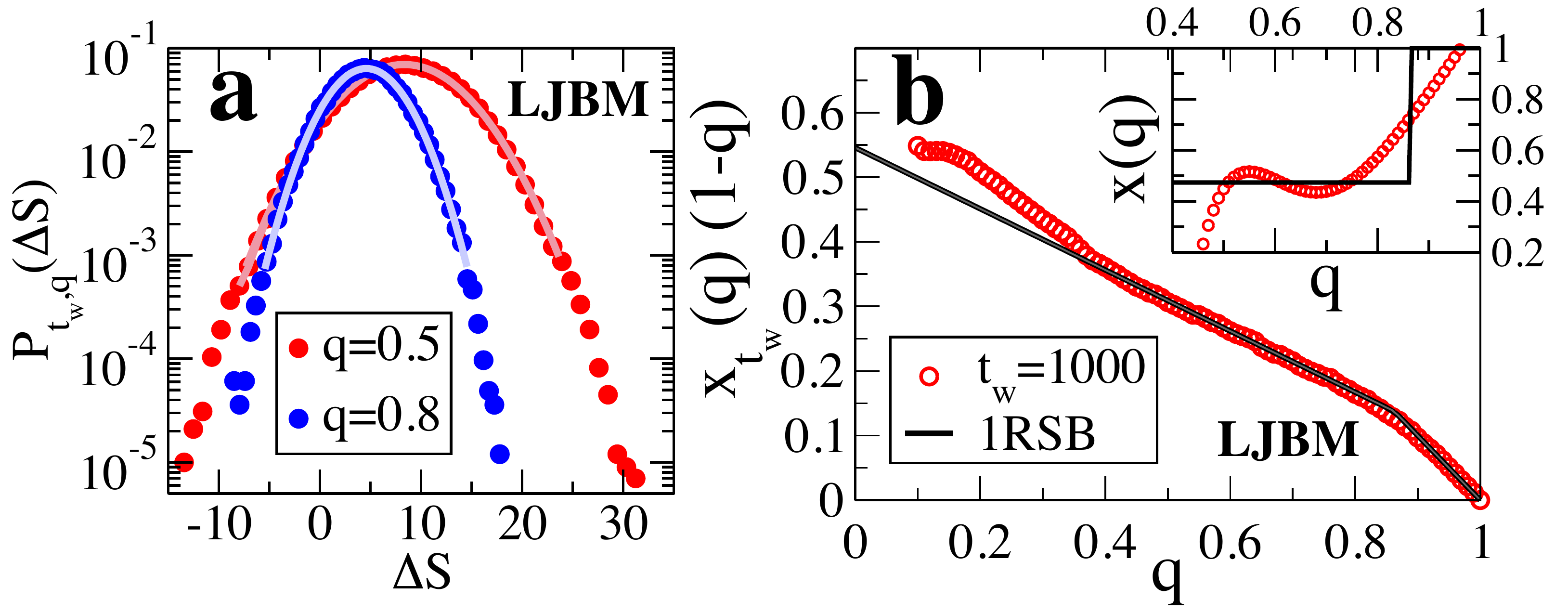} 
   \caption{{\bf Numerical test in the BMLJ model.} (a) $q$-stat distributions for $t_w=1000$. Light-colour lines are Gaussian fits.  (b) $(1-q)x_{t_w}(q)$ and $x(q)$ (inset) versus $q$ from (\ref{xqGaussian}) for $t_w=1000$ (red) compared to 1RSB with parameters taken from FD plots \cite{CriPicRit13}. Energy and length parameters of the BMLJ (\ref{eqBMLJ})
are measured in units of $\sigma_{AA}$ 
and $\epsilon_{AA}$:
$\epsilon_{BB} = 0.5$, $\epsilon_{AB} = 1.5$, 
$\sigma_{BB}=0.88$ and $\sigma_{AB} = 0.80$, 
and are taken to prevent crystallization \cite{KobAnd94}. The system has a reduced density $\rho = 1.2$
and exhibits a glass transition at $T_g  \simeq 0.435$.
Simulations parameters: $T=0.3,N=500,h=0.1,$ 20k runs.}
   \label{fig:FIG4}
\end{figure}

The $x(q)$ describes the statistics of the free energy landscape of DFS at the level of individual samples, being also applicable to systems without quenched disorder. This is shown in Fig.\ref{fig:FIG4} where we extract the $x(q)$ for the $80:20$ binary mixture of type $A$ and $B$ particles 
interacting via a Lennard-Jones pair potential (BMLJ):
\begin{equation}
 V_{\alpha\beta}(r) = 4\,\epsilon_{\alpha\beta} \left[
       \left(\frac{\sigma_{\alpha\beta}}{r}\right)^{12}
              -
       \left(\frac{\sigma_{\alpha\beta}}{r}\right)^{6}
                                             \right] 
                                             \label{eqBMLJ}
\end{equation}
where $\alpha, \beta = A, B$,
$r$ is the distance between the two particles and the parameters
  $\sigma_{\alpha\beta},\epsilon_{\alpha\beta}$ stand for the effective
  diameters and well depths between species $\alpha,\beta$. Remarkably, the $x(q)$ in the BMLJ (Fig.\ref{fig:FIG4}b, inset) shows combined features of 1RSB (plateau at $q\lesssim 0.8$) and full RSB ($q\gtrsim$ 0.8). 
%

\tcr{We have introduced a FR to extract the order parameter $q(x)$ in DFS based on measuring entropy production fluctuations at a fixed value of the overlap $q$.  Equation (\ref{eqLDF2}) and the GUH should be seen as key features of a weak ergodicity breaking scenario where $q$-stat classification of paths provides a operational definition of partially equilibrated states in rugged free energy landscapes with quasi-FDT dynamics (c.f. Eq.(\ref{QFDT})). The approach has been validated for three kinds of glassy systems being potentially applicable to other models and scenarios such as the low-temperature Gardner-like RSB phase. Initially discovered in the Ising p-spin-glass \cite{Gar85} the Gardner transition has been shown to describe the spectrum of low energy excitations in jamming systems \cite{Wya12,Cha14}. Moreover, it should be applicable to  small systems \cite{Sei12}, such as biomolecules \cite{CamAleRit17} or microspheres trapped in random media \cite{GomPetCil11}, where thermal forces induce measurable energy fluctuations. A main difficulty in experiments is the measurement of the overlap between microscopic configurations, $q=x(t)x(s)$ where $x$ stands for a reaction coordinate. Recent experimental studies \cite{Die15} have shown the way to extract such quantities by  measuring and testing the validity of the quasi-FDT in microspheres and single molecules mechanically driven to nonequilibrium steady states using steerable optical traps. Such measurements and the identification of suitable molecular systems exhibiting slow relaxational behavior reminiscent of DFS remains an exciting route to pursue in the near future. Nearly half a century ago, while mean-theory of the spin glass and replica symmetry breaking were developed, it was expected that mean-field results could ultimately be applied to finite-dimensional systems. The present work will hopefully shed light into this fundamental and still unanswered question.}

\begin{acknowledgments}
AC thanks the LPTHE for kind hospitality. FR acknowledges
support from  ICREA Academia 2013 and Spanish Research Council Grant FIS2016-80458- P
\end{acknowledgments}

\bibliographystyle{unsrt}

\end{document}